# Maximum Likelihood Channel Decoding with Quantum Annealing Machine

Naoki Ide
*Advanced Research Lab., R&D Center*
*Sony Corporation*
Tokyo, Japan
Naoki.Ide@sony.com

Tetsuya Asayama
*Advanced Research Lab., R&D Center*
*Sony Corporation*
Tokyo, Japan
Tetsuya.Asayama@sony.com

Hiroshi Ueno
*Advanced Research Lab., R&D Center*
*Sony Corporation*
Tokyo, Japan
Hiroshi.C.Ueno@sony.com

Masayuki Ohzeki
*Sigma-i Co., Ltd, Tohoku University,*
*Tokyo Tech*
Tokyo, Japan
mohzeki@sigmailab.com

*Abstract*— We formulate maximum likelihood (ML) channel decoding as a quadratic unconstraint binary optimization (QUBO) and simulate the decoding by the current commercial quantum annealing machine, D-Wave 2000Q. We prepared two implementations with Ising model formulations, generated from the generator matrix and the parity-check matrix respectively. We evaluated these implementations of ML decoding for low-density parity-check (LDPC) codes, analyzing the number of spins and connections and comparing the decoding performance with belief propagation (BP) decoding and brute-force ML decoding with classical computers. The results show that these implementations are superior to BP decoding in relatively short length codes, and while the performance in the long length codes deteriorates, the implementation from the parity-check matrix formulation still works up to 1k length with fewer spins and connections than that of the generator matrix formulation due to the sparseness of parity-check matrices of LDPC.

## I. INTRODUCTION

LDPC [1] is a well-known channel coding technology, widely adopted in many wireless communication technology fields, such as 5G and broadcasting. LDPC has adopted BP decoding as the de facto standard algorithm for its capacity achieving performance and tractable processing time. BP decoding is an algorithm that iteratively propagates the belief values between variable nodes and check nodes.

Maximum-likelihood (ML) channel decoding, known to be the theoretically best decoding method and expected to be superior to BP decoding, has not been considered as a practical method for block codes because of the exponential increase in processing time with respect to the code length.

Quantum annealing [2] is known as a kind of quantum computing specialized to solve combinatorial optimization problems, especially QUBO, by the adiabatic transition process of Ising models keeping in their ground states. The Ising model is described by the following Hamiltonian,

$$H = -\sum_{i=1}^{N}\sum_{j=1}^{N} J_{ij}\sigma_i\sigma_j - \sum_{k=1}^{N} h_k\sigma_k \qquad (1)$$

where $N$ is the number of spins, $J_{ij}$ is the coupling coefficient of the $i$-th and $j$-th spins $\sigma_i$, $\sigma_j$, and $h_k$ is the local coefficient of the $k$-th spin $\sigma_k$. In quantum annealing, combinatorial optimization problems are translated into Ising models, calculating these coefficients $J$ and $h$. After getting the coefficients, the quantum annealing starts from the initial Hamiltonian only with transversal field that realizes quantum superposition of all possible spin states and makes the system gradually transit to the Ising Hamiltonian that realizes the optimal spin combination of the given problem.

The device implementation of quantum annealing produced by D-Wave [3] has been growing its scale in the last decade. The current D-Wave machine realizes Ising models with about 2000 spins on its hardware system. However, due to restrictions on the physical layout, the maximum number of connections that can be physically combined is restricted to 6 per spin [4]. The objective of having a larger number of spins and connections can be realized by the hybrid-solver service, but it is difficult to extend the quantum tunneling effect, which makes the system avoid the local minimum, to the extended spins and connections. Although the machine is still in the development phase, recently some of the research demonstrate promising results of solving combinatorial optimization by the quantum annealing machine in finance [5] and robotics [6].

In this paper, we will show the formulation of the ML decoding into Ising models and the experimental results of the implementations of them for LDPC by the quantum annealing machine. The following sections explain two types of ML decoding with Ising model formulations from the generator matrix and the parity-check matrix, the experimental settings of channel simulation of LDPC and the results and discussion of the decoding performance of the implementations of Ising formulations compared with the BP decoding and the brute force ML decoding with classical computers.

## II. METHOD

### A. Low-density Parity-check Codes

LDPC codes can be defined by a sparse parity-check matrix. The parity-check matrix $H$ is a $K \times N$ binary matrix, where $K$ is the parity length and $N$ is the code length, that calculates the parity-check code $c$ of the bit sequence $x$ by $c = \mathrm{mod}2(Hx)$, where mod2 is a function to output the remainder of modulo 2 of input. The parity-check code $c$ must satisfy 0 to make the bit sequence $x$ belong to a LDPC code space. "Sparse" means that the parity-check matrix has few non-zero elements in each row and column. BP decoding is an algorithm that regards the parity-check matrix as a bipartite graph called Tanner graph and calculates likelihood ratio for each message bit using the sum product algorithm.

There are two types of LDPC, regular LDPC and irregular LDPC. While the regular LDPC can be defined by the parity-check matrix with fixed number of non-zero elements in each row and column, the irregular LDPC does not have such kind of regularity. In the following paragraph, we discuss the regular LDPC, which has a fixed number of non-zero elements $w_v$ in columns and $w_c$ in rows.

While LDPC codes are defined by the parity-check matrix, we need to obtain a $N \times M$ binary generator matrix $G$ to encode message bits $\boldsymbol{m}$ into code bits $\boldsymbol{x}$ by $\boldsymbol{x} = \mathrm{mod}2(G\boldsymbol{m})$. The generator matrix must satisfy $\mathrm{mod}2(HG) = 0$, and typically satisfies the form $G = [I; Q]$, where $I$ is a $M \times M$ unit matrix, $Q$ is a $(N - M) \times M$ binary matrix and $[;]$ represents concatenation of the input matrices. The encoding methods with this type of generator matrix are called systematic encoding and when we use systematic encoding, the first $M$ bits of encoded bits are equal to the message bits, which is a useful feature for decoding.

*B. Ising formulation from the generator matrix*

ML decoding of linear binary codes in communication channels with additive white Gaussian noise can be formulated by following expression,

$$\boldsymbol{m}^* = \underset{\boldsymbol{m}}{\mathrm{argmin}} \|\boldsymbol{r} - \mathrm{mod}2(G\boldsymbol{m})\|^2 \qquad (2)$$

where $\boldsymbol{m}^*$ is the decoded message of a bit sequence, $\boldsymbol{m}$ is a candidate message, $\boldsymbol{r}$ is a received signal, corresponding to an encoded message with additive white Gaussian noise, $G$ is a binary generator matrix. Starting from this expression we can obtain following objective function [7],

$$L(\boldsymbol{\sigma}) = \sum_{i=1}^{N}\left(r_i - \frac{1}{2}\right)\prod_{k=1}^{M_i} \sigma_{i_k} \qquad (3)$$

where $\boldsymbol{\sigma}$ is the spin vector translated by $\boldsymbol{\sigma} = 1 - 2\boldsymbol{m}$, $N$ is the length of codes, $M_i$ is the number of non-zero elements in $i$-th row of the generator matrix and $i_k$ is column index of the $k$-th non-zero element in the $i$-th row. This can be derived from the following relation.

$$\mathrm{mod}2\left(\sum_{j=1}^{M} g_{ij}m_j\right) = \frac{1}{2}\left(1 - \prod_{j=1}^{M}(1 - 2g_{ij}m_j)\right) \qquad (4)$$

To implement the problem into Ising form, we need to reduce the dimension of total power in (3) into quadratic form of spin variables as (1). We can reduce the dimension by introducing auxiliary spins as following way.

$$\begin{aligned}
\prod_{k=1}^{M_i} \sigma_{i_k} &= \sigma_{i_1}\sigma_{i_2}\sigma_{i_3}\ldots\sigma_{M_i} \\
&= p_{i_2}\sigma_{i_3}\ldots\sigma_{M_i} \quad (p_{i_1} = \sigma_{i_1}, p_{i_2} = p_{i_1}\sigma_{i_2}) \\
&= p_{i_m}\sigma_{i_{m+1}}\ldots\sigma_{M_i} \quad (p_{i_m} = p_{i_{m-1}}\sigma_{i_m}) \\
&= p_{i_{M_i}} \qquad \left(p_{i_{M_i}} = p_{i_{M_i-1}}\sigma_{i_{M_i}}\right) \qquad (5)
\end{aligned}$$

where $p_{i_m}$ is the product of two spins $p_{i_{m-1}}\sigma_{i_m}$. Although this replacement reduces the dimension of total power into first-order term, we need to have additional penalty terms expressing the constraints of these spin variables. The following formulation is an example of the penalty term,

$$H_{i_m} = \frac{1}{2}\left(p_{i_m}p_{i_{m-1}} + p_{i_{m-1}}\sigma_{i_m} + p_{i_m}\sigma_{i_m}\right)$$
$$+ \left(a_{i_m} + \frac{1}{2}\right)\left(2a_{i_m} - p_{i_m} - p_{i_{m-1}} - \sigma_{i_m}\right) \qquad (6)$$

where $a_{i_m}$ is an auxiliary spin that allows the penalty term to be zero when the constraints are satisfied and otherwise positive nonzero. This penalty term takes the minimum value 0 if the constraint condition of $p_{i_m} = p_{i_{m-1}}\sigma_{i_m}$ is satisfied and $a_{i_m}$ is appropriately selected, and has a positive non-zero value otherwise. This can be confirmed by listing up values of the penalty terms for 16 combinations of $(p_{i_m}, p_{i_{m-1}}, \sigma_{i_m}, a_{i_m})$. Adding these penalty terms into the objective function (3), we can obtain the modified objective function,

$$H(\boldsymbol{\sigma}, \boldsymbol{p}, \boldsymbol{a}) = \sum_{i=1}^{N}\left(\left(r_i - \frac{1}{2}\right)p_{i_{M_i}} + \sum_{m=2}^{M_i} \lambda_{i_m} H_{i_m}\right) \qquad (7)$$

where $\lambda_{i_m}$ is a positive real hyper parameter. It is straightforward to confirm that when the auxiliary spins satisfy the constraints the second term becomes zero and the objective function matches (3). Also, we can confirm that this objective function is a quadratic form of spin variables, satisfying the Ising Hamiltonian of (1). We can derive the first and second order coefficients $\boldsymbol{h}, \boldsymbol{J}$ of Ising Hamiltonian from this expression and set these coefficients to a quantum annealing system to get the spin configuration that minimizes the objective function. After getting the optimized spins $\boldsymbol{\sigma}^*$ from the quantum annealing system, we convert the resulting spins to the message bits by $\boldsymbol{m}^* = (1 - \boldsymbol{\sigma}^*)/2$.

*C. Ising formulation from the parity-check matrix*

Another formulation of ML decoding can be written in the following way.

$$\boldsymbol{x}^* = \underset{\boldsymbol{x}}{\mathrm{argmin}}[\|\boldsymbol{r} - \boldsymbol{x}\|^2 + \boldsymbol{\lambda} \cdot \mathrm{mod}2(H\boldsymbol{x})] \qquad (8)$$

where $\boldsymbol{x}^*$ is the decoded message with parity bits, $\boldsymbol{x}$ is a candidate message, $\boldsymbol{\lambda}$ is a coefficient vector and $H$ is a binary parity-check matrix. We can confirm that this form is equal to the ML decoding of (2) when the message codes satisfy parity-check constraints. Using the similar relation of (4), we can obtain the following objective function.

$$L(\boldsymbol{\sigma}) = \sum_{i=1}^{N}\left(r_i - \frac{1}{2}\right)\sigma_i + \frac{\lambda}{2}\left(K - \sum_{k=1}^{K}\prod_{m=1}^{N_k} \sigma_{k_m}\right) \qquad (9)$$

where $\boldsymbol{\sigma}$ is the spin vector translated by $\boldsymbol{\sigma} = 1 - 2\boldsymbol{x}$, $K$ is the number of the parity bits, $N_k$ is the number of non-zero elements in $k$-th row of the parity-check matrix and $k_m$ is column index of the $m$-th non-zero element in the $k$-th row. While we can set different coefficients to each parity constraint in $\boldsymbol{\lambda}$, we set them with the identical value $\lambda$. Following the formulations similar to (5) and (6), we obtain the following modified objective function,

$$H(\boldsymbol{\sigma}, \boldsymbol{p}, \boldsymbol{a}) = \sum_{i=1}^{N}\left(r_i - \frac{1}{2}\right)\sigma_i + \frac{\lambda}{2}\left(K - \sum_{k=1}^{K} p_{k_{N_k}}\right)$$
$$+ \sum_{k=1}^{K}\sum_{m=2}^{N_k} \lambda_{k_m} H_{k_m} \qquad (10)$$

where $\lambda_{k_m}$ is a positive real hyper parameter. Similar to the previous formulation of (6), this objective function also satisfies the Ising Hamiltonian (1) and we can get the optimized spins $\boldsymbol{\sigma}^*$ from the quantum annealing system. It is easy to convert the resulting spins to code bits by $x^* = (1 - \sigma^*)/2$ and we can get message bits by $m^* = x^*_{1...M}$ for systematic encoding.

*D. Estimation of the number of spins and connections.*

To implement these Ising models into the quantum annealing machines, there are the limitations of the maximum number of spins and connections between two spins. We estimate them for each formulation as following.

In the case of the generator matrix formulation, we can estimate the number of spins by counting non-zero elements in each row of the generator matrix. Assuming systematic encoding, the generator matrix satisfies the form $G = [I; Q]$. In each row of the unit matrix $I$, there are no additional spins to reduce the dimension of the formulation (3). On the other hand, in each row of the matrix $Q$, the $(N - M) \times M$ matrix, the number of non-zero elements is $M/2$ on average. As formulated previously, we need two additional spins for each nonzero element in each row. Therefore, we need about $2 \times (N - M) \times (M/2 - 1)$ additional spins and $M + (N - M) \times (M - 2)$ total spins on average. We can also estimate the number of connections by counting the non-zero elements in each column of the generator matrices. Since the connection terms appear in only the penalty terms of (5), we need to count the non-zero elements in $Q$. Since the number of non-zero elements in each column in $Q$ is $(N - M)/2$ on average and each spin from the message bit has 3 connections in (5), we can estimate the number of connections for each spin from the message bit as $3(N - M)/2$ on average. We can also estimate the number of connections for the auxiliary spins from (5) as 3 or 4 and therefore the maximum number of the connection can be estimated as $3(N - M)/2$ on average.

In the case of the parity-check matrix formulation, we can estimate the number of spins by counting non-zero elements in each row of the parity-check matrix. Assuming Gallager's construction [1], the $K \times N$ parity-check matrix has $w_r$ non-zero elements in each row and $w_c$ non-zero elements in each column. The Ising formulation with the parity-check matrix needs two auxiliary spins for each non-zero element in each row. Therefore, the number of the additional spins is $2K(w_r - 1)$ and the total number of spins becomes $N + 2K(w_r - 1)$. The number of connections between spins can be estimated from the number of non-zero elements in each column of the parity-check matrix. From the similar estimation as mentioned above, we can estimate the number of connections for each spin from the message bit as $3w_c$ and that of the auxiliary spins as 3 or 4. Therefore the maximum number of the connections can be estimated as $3w_c$. This estimation implies that the numbers of spins and connections are less likely to increase in the parity-check matrix formulation than in the generator matrix formulation and the longer code can be decoded by the parity-check matrix formulation under the quantum annealing system with limited spins and connections.

## III. EXPERIMENT

*A. Experimental setup*

We performed numerical simulations of regular LDPC codes in additive white Gaussian noise channel to estimate the performance of the implementations of Ising formulations.

We used a python library to generate parity-check matrices and generator matrices of regular LDPC [8]. From this library, parity-check matrices can be generated with given parameters $n$, the code length, $w_r$, the number of non-zero elements in each row of the parity-check matrix and $w_c$, the number of non-zero elements in each column of the parity-check matrix. Also, the generator matrix is generated by the parity-check matrix with reducing its matrix rank properly.

We prepared the communication channel simulation with BP decoding and ML decoding. We had two proposed methods, the Ising formulation of ML decoding from the generator matrix, and that from the parity-check matrix. The hyper parameters $\lambda_{i_m}$ of the generator matrix formulation were all set to the same value 0.5. The parameter $\lambda$ of the parity-check matrix formulation was set to 0.3. The parameters $\lambda_{k_m}$ were all set to the same value 0.5. These parameters were determined by simple grid search in increments of 0.1. In each formulation, we used the quantum annealing machine D-Wave 2000Q through D-Wave's Hybrid Sampler API. Also we had three reference methods, a brute force ML decoding, the BP decoding, and a simple threshold decoding. We used the BP decoding with the number of iteration fixed at 20.

We constructed an evaluation system that evaluates bit error rate (BER) in the noisy channel with a variety of energy per bit to noise density ratio ($E_b/N_0$) settings. For the LDPC settings, we used settings in which the code lengths were doubled from 16 to 1024. The number of non-zero elements in each row of the parity-check matrix was fixed at 8, and that in each column was fixed at 4. At these settings, the coding rates are about 0.5 to 0.7.

We evaluated the BER of these encoding and decoding settings under various noise conditions. The magnitude of the noise was calculated from $E_b/N_0$. In our evaluation, $E_b/N_0$ was changed from 0 to 5 or 8, and the BER was measured in each setting. The BER evaluation was performed on a fixed number of blocks of 100 blocks or more. Specifically, the BER was evaluated with 400 blocks when the code length was 16, 300 blocks, when the code length was 32, 200 blocks, when the code length was 64, and 100 blocks otherwise.

*B. Comparison of the number of spins and connections*

The following table shows the number of spins and the maximum number of connections in each LDPC setting.

TABLE I.  ISING MODEL ANALYSIS

| Code length | Coding rate | Generator Form | | Parity-check Form | |
|---|---|---|---|---|---|
| | | # spins | max # connection | # spins | max # connection |
| 16 | 0.69 | 45 | 9 | 112 | 12 |
| 32 | 0.59 | 169 | 26 | 224 | 12 |
| 64 | 0.55 | 733 | 51 | 448 | 12 |
| 128 | 0.52 | 3085 | 105 | 896 | 12 |
| 256 | 0.51 | 12665 | 186 | 1792 | 12 |
| 512 | 0.51 | 51657 | 381 | 3584 | 12 |
| 1024 | 0.50 | 212317 | 717 | 7168 | 12 |

As shown in Table I, the number of spins drastically increases in the generator matrix formulation as the code length increases, but not as much in the parity-check matrix formulation. Also, the maximum number of spin connections gradually increases in the generator matrix formulation, but is constant in the parity-check matrix formulation. These results match the estimation in the previous section.

## C. Comparison in relatively short length LDPC

As the examples of LDPC having a short code length, the comparisons of the decoding methods in case of the code length 32 and 64 are shown in Fig. 1.

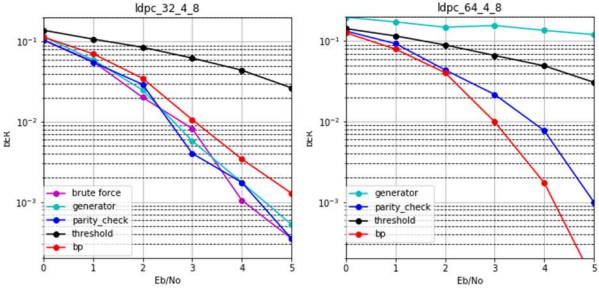

Figure 1. $E_b/N_0$-BER curve results in relatively short length LDPC. Left: Code length 32, Right: Code length 64, both the parameter $w_r, w_c$, the number of non-zero elements in each row and column respectively are fixed 8 and 4.

As shown in the left panel of Fig. 1, in case of the code length 32, the implementations of the ML decoding realized higher performance than BP decoding at the same $E_b/N_0$. On the other hand, it is difficult to find a clear difference between the ML decoding implementations. This result was similar for the length 16.

On the other hand, as shown in the right panel of Fig. 1, remarkable differences appeared between the ML decoding implementations when the code length was 64.

First, our implementation of the brute force ML decoding could not be measured because it took too much time for even single decoding. This is because the decoding time increases in exponential order with respect to the message length.

Subsequently, in the Ising implementation with the generator matrix formulation, although the decoding process operated in about 10 seconds each time, the BER deteriorated significantly and the result was upper than the simple threshold decoding. The reason for this result is in the embedding in the sparsely connected graph, on which the quantum annealing machine solves the effective model imitating the original problem. The number of spins and the maximum number of connections for a code length of 64 are 733 and 51 respectively and it exceeds the limitation of the connections. The lack of connections is compensated by the additional ones of physically-connected spins, often termed as spin chain. The performance of the quantum annealing machine strongly depends on the length of spin chains and structure of the embedding on the sparsely connected graph. Although the current quantum annealing machine can avoid the sparseness problem with the hybrid approach (D-Wave hybrid) of the existing computation and the quantum annealing, we here focus on the performance of the quantum annealing itself, which is not yet realized as a perfect full connection.

On the other hand, in the parity-check matrix formulation, while the decoding process also operated in about 10 seconds at a time, there was no noticeable deterioration in the BER as seen in the generator matrix formulation. In the parity-check matrix formulation, the number of spins and connections for the code length of 64 are 448 and 12, respectively. Compared to the case of the generator matrix, the number of the necessary connection was smaller. From this observation, the performance of the quantum annealing strongly depends on the embedding. Thus, there is a remarkable difference in performance between the generator matrix formulation and the parity-check matrix formulation.

Compared with the BP decoding, the parity-check matrix formulation is slightly degraded unlike the code length of 32. Since ML decoding is theoretically the best decoding, at least the same performance as the BP decoding is expected, but the result was different otherwise.

There are several possible reasons for this result. One of them is related to the proposed formulation. For example, we determined the hyper parameters in 0.1-step grid searches, but the BER was quite sensitive to these parameters. Another possible reason is due to the quantum annealing machine itself, such as the embedding of the problem, the interaction with the environment and the resolution of coefficients. As for the embedding, as discussed before, the modification from the original problem by use of the physically-connected spins degrade the performance of the quantum annealing machine for solving the original problem. The effect strongly depends on the number of connections beyond the limitation of the quantum annealing machine. The interaction with the environment hampers the ideal process of the quantum annealing, which was originally the method in the isolated quantum system. The actual quantum machine is open quantum system and the environment effect spoils the quantum tunneling. As for the resolution of the coefficients, if it is not high enough, the ground state becomes quite different from the original ground state. If the ground state is different, the result will be different from the correct answer even if the machine can reach the ground state.

## D. Comparison in relatively long length LDPC

As an example of LDPC having relatively long code length, the comparison of the decoding methods in case of the code length 1024 is shown in Fig. 2 below.

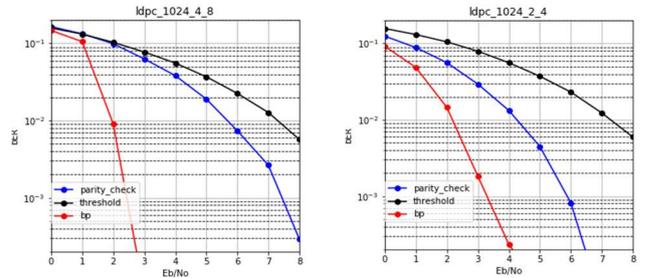

Figure 2. $E_b/N_0$-BER curve results of LDPC with code length 1024. Left: The parameters $w_r$ and $w_c$ are fixed 8 and 4 respectively. Right: $w_r$ and $w_c$ are fixed 4 and 2 respectively.

As described above, the generator matrix formulation is deteriorated significantly due to the increase in the number of connections, so only the results of the parity-check matrix formulation is shown. When the code length becomes longer, while the steep waterfall region appears in the BP decoding in BER curves, the Ising implementations did not show such results. This difference gradually increases as the code length changes from 64 to 1024. It is estimated that as the code length becomes longer, the problems of local minimum and resolution become more prominent. We have also the result of LDPC with the other setting. In this setting, we set $n$ as 1024, $w_r$ as 4 and $w_c$ as 2 which corresponds to the number of spins and connections as 3072 and 6, expecting the effect to reduce the maximum number of connections under the limitation of the machine. The result shows the improvement compared with the original setting.

## E. Computation time

Fig. 3 shows the time profile of the computation for the parity-check matrix formulation.

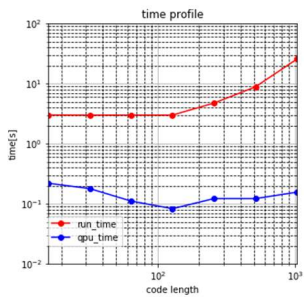

Figure 3. Time profile regarding to code length. The runtime and QPU access time are averaged after being obtained from D-Wave's sampler API.

From the D-Wave's hybrid API, we got the information of runtime, charge time and QPU access time. As shown in Fig. 3, the runtime is about 3 to 30 seconds, which is too long to use as a communication decoding technique. On the other hand, the QPU access time included in the runtime is below 300 ms, which is permissible as a channel decoding time. The time required for quantum annealing to be the optimal solution said to increases exponentially with respect to the number of spins in the worst case [9]. However, in this measurement, the quantum annealing process was below 200 ms, and the dominant part of the computation was CPU processing. Since the convergence time is automatically processed by the machine, it cannot be denied that the convergence time may not be sufficient to find the optimal solution. However, if the solution can be obtained in about 300 ms in the Ising implementations, it can be considered for channel decoding. Of course, it is not desirable to use such a large quantum annealing machine inside mobile devices or to communicate with the cloud service to decode the channel codes each time. However, fortunately, there are various Ising machines that can be used as substitutes of the quantum annealing machine and can be realized only with existing transistors [10, 11, 12].

## IV. DISCUSSION

Discussion of channel decoding using Ising model has been around for relatively long period. The straightforward formulation of the decoding problem into Ising model yields to the many body interactions among the Ising variables [7]. Our main contribution is to show how to implement the ML decoding for LDPC with realistic code length under the constraints of the current real quantum annealing machines. Although the embedding problem remains, we purely test the performance of the quantum annealing machine. Apart from the Ising formulation, there are various studies of approximate ML decoding, such as gradient-based method [13]. However, a different perspective from the Ising formulation contributes to the development of coding theory, quantum annealing, and Ising machines.

One of the applications of this technology is the ML decoding other codes than LDPC ones. Reed Solomon codes, BCH codes and Hamming codes, that have algebraic decoding, can be considered. For realizing the decoding of these codes, first, the generator matrix or the parity-check matrix should be represented by binary matrices. Furthermore, it would be effective to minimize the number of non-zero elements of the parity-check matrix, especially in the column direction which is related to the number of connection, as much as possible by applying basic transformation regarding the rows of matrix.

It is also expected to be applied in terms of the performance evaluation of Ising machines. While the brute force ML decoding can give the optimum solution, its operation is intractable due to the combinatorial explosion. On the other hand, BP decoding works with even a long code. Although BP decoding is not the optimum, it has quite high capacity achieving performance. Therefore, the BER of BP can be regarded as a bound of the BER of ML decoding and treated as a milestone for improving the performance of Ising machines. Fortunately, the implementation of the ML decoding by Ising formulation is extremely simple. In addition, the library of LDPC is relatively substantial, and the BP decoding can be easily used. Thus, this research can contribute as one of the milestones for improvement of Ising machines.

## V. CONCLUSION

In this work, we performed the Ising formulations for ML decoding from the generator matrix and the parity-check matrix. We showed that in the current quantum annealing machines, the parity-check matrix formulation can be implemented more easily if the matrix is sparse, while the generator matrix formulation is difficult to implement due to drastically increases of the number of connections. Based on these formulations, we implemented the ML decoding of LDPC in a quantum annealing machine and compared the performance of these implementations with BP decoding. As a result, we showed that the Ising implementation of ML decoding has higher performance than BP decoding in relatively short code lengths, while it does not reach the performance of the BP decoding with relatively long codes.

## ACKNOWLEDGMENT

We appreciate Sigma-i Co., Ltd. for their co-operation.